\documentclass[11pt,a4paper]{article}

\usepackage{jheppub}
\usepackage{bbold}

\makeatletter
\def\@fpheader{Prepared for submission to JCAP}
\makeatother

\makeatletter
\DeclareRobustCommand{\rcite}[1]{%
  \rcite@aux#1,\@nil{#1}%
}
\def\rcite@aux#1,#2\@nil#3{%
  \if\relax#2\relax
    Ref.~\cite{#3}%
  \else
    Refs.~\cite{#3}%
  \fi
}
\makeatother

\title{Scalar instabilities in bimetric gravity: The Vainshtein mechanism and structure formation}

\author[a,b]{E. M{\"o}rtsell}
\author[a,b]{J. Enander}
\affiliation[a]{Oskar Klein Centre, Stockholm University,\\Albanova University Center\\ 106 91 Stockholm, Sweden}
\affiliation[b]{Department of Physics, Stockholm University\\
AlbaNova University Center\\ 106 91 Stockholm, Sweden}
\emailAdd{edvard@fysik.su.se}
\emailAdd{enander@fysik.su.se}
\abstract{We investigate the observational consequences of scalar instabilities in bimetric theory, under the assumption that the Vainshtein mechanism restores general relativity within a certain distance from gravitational sources. We argue that early time instabilities have a negligible impact on observed structures.
Assuming that the instabilities affect sub-horizon density fluctuations, we constrain the redshift, $z_i$, below which instabilities are ruled out. For the ``minimal'' $\beta_1$-model, observational constraints are close to the theoretical expectations of $z_i\approx 0.5$, potentially allowing the model to be ruled in or out with a more detailed study, possibly including secondary cosmic microwave background constraints.}

\keywords{modified gravity, bigravity, massive gravity, dark energy, cosmic acceleration}

\begin{document}

\maketitle
 
\section{Introduction}

General relativity is the unique theory of a massless spin-2 field interacting with matter. While it has been tested to great precision within the solar system, its application on the level of galaxies and beyond points towards the presence of major unknown components of the universe---dark energy and dark matter---which have so far only been probed gravitationally. From a theoretical point of view, the observed value of the cosmological constant seems to require a high degree of fine-tuning in order to be reconciled with the quantum properties of the vacuum. Adding to this observational conundrum is the fact that general relativity is non-renormalizable as an effective field theory. These observational and theoretical problems have sparked a great deal of interest in possible extensions of general relativity in order to fully understand the nature of gravitational interactions.

Following the classification of the particles in the Standard Model according to their mass and spin, a natural extension of general relativity is to introduce another spin-2 field into the gravitational sector. This gives rise to both massive and massless spin-2 interactions. Such proposed interactions have a long history \cite{Fierz:1939ix,Isham:1971gm}, but finding the adequate theoretical formulation has proven difficult due to the presence of ghost modes in the spectrum \cite{Boulware:1973my}. Recently, a formulation, dubbed the Hassan-Rosen theory, with two interacting spin-2 fields was found that does not contain a ghost \cite{2012JHEP...02..126H,2012PhRvL.108d1101H}. This formulation contains the so-called de Rham-Gabadadze-Tolley theory  \cite{2010PhRvD..82d4020D,2011PhRvL.106w1101D} in the limit when one of the fields becomes non-dynamical. That these theories were ghost free was proven in \cite{2012PhRvL.108d1101H,Hassan:2011hr,Hassan:2011tf,Hassan:2012qv}.

The background cosmology of the Hassan-Rosen theory has been extensively studied \cite{vonStrauss:2011mq, Volkov:2011an, Comelli:2011zm, Akrami2013JHEP...03..099A, Akrami:2013pna, Nersisyan:2015oha}. It was shown, however, in Ref.~\cite{Comelli:2012db} that certain instabilities could arise in the scalar sector on a perturbative level. The scalar sector was further studied in \cite{Khosravi:2012rk,Berg:2012kn,Fasiello:2013woa,Konnig:2013gxa,Konnig:2014dna,Comelli:2014bqa,DeFelice:2014nja,Solomon:2014dua,Konnig:2014xva,Lagos:2014lca,Enander:2015vja,Konnig:2015lfa,Aoki:2015xqa}. In particular, Ref.~\cite{Konnig:2014xva} gave conditions for the existence of the scalar instabilities in terms of the parameters of the Hassan-Rosen theory. 

The instabilities are also present in the tensor sector, and have been studied in Refs.~\cite{Cusin:2014psa,Johnson:2015tfa,Amendola:2015tua,Fasiello:2015csa}. In all of these studies, it was assumed that matter only couples to one of the fields of the theory. The possibility of coupling matter to both fields was explored in Refs.~\cite{2015GReGr..47.1838A,deRham:2014naa,deRham:2014fha,Enander:2014xga,Schmidt-May:2014xla,Comelli:2015pua,Gumrukcuoglu:2015nua,Akrami:2013ffa,Solomon:2014iwa,
Gumrukcuoglu:2014xba,Heisenberg:2015iqa}. The inclusion of other matter couplings does not seem to alleviate the problem of the instabilities \cite{Comelli:2015pua,Gumrukcuoglu:2015nua}.

In this note, we address the question of whether scalar gradient instabilities automatically 
disqualifies bimetric theory or only make the theory extremely complicated 
phenomenologically since linear perturbation theory is not valid. Ref.~\cite{Akrami:2015qga} showed that the instabilities can be removed by taking the general relativity limit of the Hassan-Rosen theory, in which the effective Compton wavelength of the massive graviton goes to zero. On qualitative and semi-quantitative grounds,
we argue for another alternative, namely that under the assumption of a Vainshtein mechanism, scalar instabilities have a negligible impact on structure formation, at least at high redshifts, and that linear structure formation theory can be used in the same regimes as in GR. 

We will use units where $G=c=1$, where $c$ here refers to the velocity of light. In the remainder of the paper, however, it will denote the relation between 
the asymptotic background metrics, as described below.

\section{The Vainshtein mechanism}

When expanding the metric fields around a flat background, higher order perturbative terms will become important inside the so-called Vainshtein radius, $r_V$ \cite{1972PhLB...39..393V}. This regime does not necessarily correspond to the case where all the metric perturbations, or even density fluctuations, become of order unity. The existence of the Vainshtein mechanism for static and spherically symmetric (SSS) systems in the context of bigravity was shown in Refs.~\cite{Volkov:2012wp,Babichev:2013pfa}. An explicit expression of the second order term and the Vainshtein radius that they implied in terms of the parameters of the Hassan-Rosen theory was presented in Ref.~\cite{2013JHEP...10..031E} (see also Ref.~\cite{Enander:2015kda}).

Due to the complexity of the equations, the Vainshtein mechanism and possible recovery of GR at 
small distances from the source have not been studied analytically for non-spherically symmetric and non-static 
systems in bimetric theory. However, in the  
following, we will assume that also for time varying density fluctuations, the Vainshtein mechanism 
will become effective at a distance similar to the corresponding Vainshtein radius for the SSS-case and
that GR is recovered at $r\ll r_{V}$. Justification of this assumption can be found for the DGP and 
cubic Galileon model in Refs.~\cite{2009PhRvD..80d3001S,Winther:2015pta,Brito:2014ifa}, see also 
Refs.~\cite{2013PhRvL.110p1101L,2015JCAP...02..034B} for tests of the quasi-static approximation in other modified gravity models. However, for bimetric theory, this has not been shown explicitly and should only be 
regarded as an assumption at this point.

We can identify the Vainshtein radius
for the SSS case using the following diagonal forms for $g_{\mu\nu}$ and $f_{\mu\nu}$ (following the notation of Ref.~\cite{Volkov:2012wp}:
\begin{eqnarray}\label{eq:metricansatz}
ds_{g}^{2}&=&-F^{2}dt^{2}+B^{-2}dr^{2}+R^{2}d\Omega^{2},\\
ds_{f}^{2}&=&-p^{2}dt^{2}+b^{2}dr^{2}+U^{2}d\Omega^{2}.
\end{eqnarray}
We perturb the metrics around flat space ($\bar{g}_{\mu\nu}=\eta_{\mu\nu}$ and $\bar{f}_{\mu\nu}=c^2\eta_{\mu\nu}$) where, in order to have flat backgrounds, we put 
\begin{eqnarray}
\beta_{0}+3\beta_{1}c+3\beta_{2}c^{2}+\beta_{3}c^{3}&=&0,\\
\frac{\beta_{1}}{c}+3\beta_{2}+3\beta_{3}c+\beta_{4}c^2&=&0.
\end{eqnarray}
Note that if we instead of zero set the right hand sides of the above equations to $3H_0^2/m^2$, we obtain
de Sitter solutions on the background level. Equating first and second order perturbation terms, we identify the Vainshtein radius as 
\begin{equation}\label{eq:rVX}
(r^X_{V})^3=q_X(\beta_i,c)\frac{M}{m^{2}},
\end{equation}
for the metric fields $X=\{F,B,R,p,b,U\}$. Here $M$ is the mass of the source as measured by an observer at infinity and $m$ the mass scale of the theory ({\it not} equal to the effective graviton mass). The function $q_X$ depends on the parameters of the theory and is generally of order unity when $\beta_i\sim c\sim1$. As an example, it has the following form for the $p$-field:
\begin{equation}
q_p =\frac{c\left(\beta_{1}+3c\beta_{2}+2c^{2}\beta_{3}\right)}{6\left(\beta_{1}+2c\beta_{2}+c^{2}\beta_{3}\right)^{2}}.
\end{equation}
In the following we will denote with $r_V$ the largest of the Vainshtein radii (thus setting the scale at which higher order terms become important), and write
\begin{equation}\label{eq:rV}
r_{V}^{3}=\frac{M}{m^{2}}q=M\lambda^2 q,
\end{equation}
where $\lambda=m^{-1}$ and 
\begin{equation}\label{eq:rVX}
q\equiv\max [q_X(\beta_i,c)].
\end{equation}
As will be discussed in Sec.~\ref{sec:b1}, the value of $q$ will depend not only on the $\beta_i$'s and $c$, but also on the specific metric ansatz used for 
$g_{\mu\nu}$ and $f_{\mu\nu}$.

Numerically, we can write the Vainshtein radius as
\begin{equation}
r_{V}\approx0.08\left(\frac{q}{h^2}\right)^{1/3}\left[\frac{M}{M_{\odot}}\left(\frac{\lambda}{r_{H}}\right)^{2}\right]^{1/3}\,{\rm kpc},
\end{equation}
where $r_{H}=H_0^{-1}$ is the Hubble radius and $h\equiv H_0/(100\,{\rm km/s/Mpc})$. Setting $\lambda=r_{H}$ and
$M=10^{12}M_{\odot}$, corresponding to a rather
massive galaxy, we get $r_{V}\sim (q/h^2)^{1/3}\,{\rm Mpc}$. Increasing the mass by a factor of a thousand (to
obtain a heavy cluster), we increase the Vainshtein radius by a factor of
ten. We thus get a radius that naturally adapts to the size of the
object at hand. This will be of interest when studying the dynamics
of galaxies and clusters of galaxies \cite{Enander:2015kda,Koyama:2015oma}. We emphasize again that the Vainshtein radius does not correspond to the radius where {\it all} metric perturbations or density fluctuations become large. In fact, the gravitational potential $\Phi=F^2 -1\sim M/r$ is of order $\Phi\sim (M/\lambda^2)^{2/3}=(M/r_H)^{2/3}$ at $r=r_V$, which for the Sun gives $\Phi (r_V)\sim 10^{-15}$ and for a galaxy with mass $M=10^{12}M_{\odot}$, $\Phi (r_V)\sim 10^{-7}$.

\section{Densities}

The average density within the Vainshtein radius of a spherically symmetric matter collection of mass $M(r_V)$ is given by
\begin{equation}
\rho_V=\frac{M}{(4\pi r_{V}^{3}/3)}=\frac{3}{4\pi q \lambda^2},
\end{equation}
which is independent of the mass of the object. We note that this only holds for Vainshtein radii with a cubic form as in Eq.~\ref{eq:rV}. For a quintic Vainshtein radius, $r_V^5\propto M\lambda^4$, the average density within $r_V$ decreases with mass, i.e. we will recover GR to high precision for small mass fluctuations and small scales but get increasing deviations for large mass fluctuations and large scales. For Vainshtein radii of lower order than cubic, we get large deviations from GR for small mass fluctuations, for which we know observationally that any deviation from GR has to be very small. 

In the case of $\lambda=r_{H}=H_0^{-1}$,
\begin{equation}
\rho_{V}=\frac{3H_0^2}{4\pi q}=\frac{2}{q}\rho^0_{\rm crit},
\end{equation}
where
\begin{equation}
\rho_{{\rm crit}}^{0}=\frac{3H_{0}^{2}}{8\pi}\approx2.8\cdot10^{11}\, h^{2}M_{\odot}{\rm Mpc}^{-3},
\end{equation}
is the critical density of the Universe today. We thus expect the Vainshtein radius to become important for any density fluctuation larger than the critical density; a very low density indeed.

Since what is relevant for the potential fluctuations are the relative
density fluctuations $\delta\rho_m$, defined through
\begin{equation}
\frac{\delta\rho_{m}}{\bar{\rho}_{m}}\equiv\frac{\rho_{m}-\bar{\rho}_{m}}{\bar{\rho}_{m}}\equiv\delta_m,
\end{equation}
we see that in order to be within the regime where the Vainshtein mechanism is active
today, that is for $\delta\rho_{m}\gtrsim\rho_{V}=(2/q)\rho_{{\rm crit}}^{0}$, 
we need $\delta_m\gtrsim 2/(\Omega_m q)$,  
where $\Omega_m\equiv\bar{\rho}^0_{m}/\rho_{{\rm crit}}^0$.
However, since the average density increases with decreasing scale factor, at scale factor $a$,  $\bar{\rho}_{m}=\Omega_m\rho_{{\rm crit}}^{0}a^{3}$,
the density contrasts that we expect to be protected
by the Vainshtein mechanism are
\begin{equation}\label{eq:deltaV}
\delta_{m}(a)\gtrsim\delta_{V}(a)\equiv \frac{\rho_V}{\bar{\rho}_{m}}=\frac{2 a^3}{q \Omega_m}.
\end{equation}

\section{Structure formation}\label{sec:sf}

The standard picture in GR of structure formation is the following: We start out with some initial
spectrum of perturbations. These will grow according to standard
GR theory where the growth depends on whether the perturbations are sub- or super-horizon,
whether they have pressure (stabilizing them on small scales) and on the background
expansion. In the following, we only consider scalar fluctuations in pressure less
matter. It is convenient to express density fluctuations, $\delta_m$, in terms of the amplitudes of the 
Fourier components, $\delta_k$, defined by
\begin{equation}\label{eq:deltak}
\delta_k\equiv \frac{1}{V}\int \delta_m(\bar x)\exp(-i\bar k\cdot\bar x)d^3x,
\end{equation}
since they evolve independently of each other, in the linear regime. Similarly, we denote the amplitudes of the 
Fourier components for the gravitational potential $\Phi$ by $\Phi_k$.
Defining $P(k)\equiv\left\langle |\delta_{k}|^{2}\right\rangle$, where the average is over all directions, 
inflation generally generates initial conditions $P_{i}(k)\propto k^n$, where $n\approx 1$. 
Since $\delta_k\propto k^2\Phi_k$, this corresponds to $P_{i}(\Phi_k)\propto k^{n-4}\sim k^{-3}$. The potential
$\Phi_k$ will stay roughly constant outside the horizon, preserving the shape of $P(\Phi_k)$. When a Fourier mode re-enters
the horizon at $k=Ha$, 
\begin{equation}
\delta_k\approx-\frac{2}{3}\frac{k^2}{H^2a^2}\Phi_k\approx-\frac{2}{3}\Phi_k={\rm const.}
\end{equation}
That is, as perturbations enter the horizon, they will be described by $P_{H}(k)\propto k^{n-4}\sim k^{-3}$.\footnote{The power spectrum $P_H(k)$ is defined as the power at each $k$ at the time when the corresponding Fourier mode re-enters the horizon at $k=Ha$. That is, it is defined at different epochs for different scales.}
For a  mode that enters the horizon during matter domination, this will take place
at $k_{m}=H(a_{m})a_{m}\propto a_{m}^{-1/2}$. After the entry, the
mode will grow proportional to the scale factor, $\delta_k\propto a$ and the shape of the power spectrum
will be modified
\begin{equation}
P(k)=P_{i}(k)\frac{a^{2}}{a_{m}^{2}}\propto k^{-3}\cdot k^{4}=k.
\end{equation}
Modes that
enter earlier will only grow logarithmically, as long as radiation dominates.
We thus expect the shape of the power spectrum for these mode to be given by
\begin{equation}
P(k)=P_{i}(k)\log(a_{rm}/a_{r})\propto k^{-3}(\log k)^2\sim k^{-3},
\end{equation}
where $a_{rm}$ denotes the scale factor at radiation-matter equality. 
Given $P(k)$, we can calculate the expected variance on a given physical scale, $R$, as
\begin{equation}\label{eq:gaussfilter}
\sigma^2_R\equiv \left\langle |\delta_{m}(R)|^{2}\right\rangle=\frac{1}{2\pi^2}\int_0^\infty P(k)\exp(-k^2R^2)k^2dk,
\end{equation}
where we have used a Gaussian filter to single out the Fourier modes contributing to the physical fluctuations. 
For a given scale $R$, a physical fluctuation $\delta_m(R)$ receives contributions from all $\delta_k$ where $k\lesssim 1/R$.
Note also that observationally, $\delta_m\lesssim 1$ for $R\gtrsim 50$ Mpc whereas $\delta_k\gtrsim 1$ for all $k^{-1}\gtrsim 0.5\,{\rm Mpc}$.

\subsection{Instabilities}

Imagine now that the theory of gravity at hand exhibits a gradient instability which
causes very fast, typically exponential, growth of some (Fourier) modes, e.g. up to a scale factor $a_i$,  
if the perturbation is within the cosmological horizon.
Given an initial spectrum of perturbations, 
from Eq.~\ref{eq:deltaV} all (physical) fluctuations
$\delta_m\gtrsim\delta_{V}\sim 2 a_i^3/(\Omega_m q)$ are protected from gradient instabilities by the Vainshtein
mechanism, but all (sub-horizon) fluctuations with $\delta_m<\delta_{V}(a_i)$ will quickly grow to $\delta_m\sim\delta_{V}$. 
Once $\delta_m\sim\delta_{V}$, the evolution of the fluctuation would be equal to that of a fluctuation in GR, i.e. $\delta_m\sim$ constant during radiation domination and $\delta_m\propto a$ during matter domination. However, with the instability present, the perturbation will constantly adjust to the value of the Vainshtein density, and 
\begin{equation}\label{eq:dmai}
\delta_{m}(a_i)\approx\delta_{V}(a_i)\approx\frac{2 a_i^3}{\Omega_m q}.
\end{equation}
That is, fluctuations will grow like $\delta_m\propto a_i^3$, na{\"i}vely in large contrast to the observed growth of structure.
However, this predicted growth only holds for fluctuations that initially (when entering inside the horizon) have $\delta_m\lesssim\delta_{V}$, and Eq.~\ref{eq:dmai} thus only gives a lower limit on the amplitude of fluctuations as long as the instability is present. We next investigate the observational consequences of such an increased growth of the smallest fluctuation amplitudes, as a function of the scale factor $a_i$ up to which the instability is present.

\subsection{Radiation domination}

At the time of photon-baryon decoupling at $z\approx 1100$, we have $\delta_{V}\approx 5\cdot 10^{-9}/q$ and at matter-radiation equality at $z\approx 3600$, we have $\delta_{V}\approx 1\cdot 10^{-10}/q$.
Let $k_{rm}$ correspond to the $k$-value that crosses the horizon at $a_{rm}$.\footnote{We define horizon crossing as the point where $k=Ha$.} Since 
\begin{equation}
\frac{H_{rm}}{H_{0}}=\frac{\sqrt{2\Omega_{m}}}{a_{rm}^{3/2}}\,\Longrightarrow\,\frac{k_{rm}}{H_{0}}=\frac{H_{rm}a_{rm}}{H_{0}}=\sqrt{\frac{2\Omega_{m}}{a_{rm}}}\approx\sqrt{0.6\cdot3600}\approx46,
\end{equation}
i.e. $k_{rm}\approx 0.015\,({\rm Mpc}/h)^{-1}$. That is, the instability will upscale all fluctuations with 
$k>k_{rm}$ to have a minimal value of $\delta_m\approx 1\cdot 10^{-10}/q$ at radiation-matter equality (that will grow to $\delta_m\approx 5\cdot 10^{-7}/q$ today) whereas fluctuations with larger amplitude and/or size are not affected by the instabilities. Thus, the effect of an instability on sub-horizon matter perturbations during radiation domination is that for all 
scales smaller than approximately 100 Mpc, we will have a minimum amplitude of fluctuations of 
$\delta_m\approx 5\cdot 10^{-7}/q$. Using the Planck normalization of the amplitude of fluctuations at scale $R=8\,{\rm Mpc}/h$ \cite{2015arXiv150201589P}, and Eq.~\ref{eq:gaussfilter}, the average fluctuations at scales $R\lesssim 100$ Mpc have $\delta_m\gtrsim 0.3$. 
Thus, we can already anticipate that instabilities during radiation domination will have a negligible impact on the power spectrum. 

\subsection{Matter domination}
If the gradient instability is present up to a scale factor $a_i>a_{rm}$, it will set a minimum value of sub-horizon perturbations at $a_i$ of 
$\delta_{m}(a_i)\gtrsim 2 a_i ^3/(\Omega_m q)$. Since during matter domination, in GR $\delta_m$ grows proportional to the scale factor, this minimal amplitude will subsequently grow to a value today of 
\begin{equation}
\delta_{m}=\frac{a_0}{a_i}\frac{2 a_i^3}{\Omega_m q}=\frac{2 a_i^2}{\Omega_m q}\equiv\delta_{\rm min}(a_i),
\end{equation}
where $\delta_{\rm min}(a_i)$ is defined as the minimum perturbation value when an instability is present up to a scale factor $a_i>a_{rm}$. Here, and in the following, we have neglected the decrease in the growth rate when dark energy becomes dynamically important. 

When constraining the maximum scale factor at which instabilities can be present, we need to take into account the fact that the observed $\delta_m$ depends on the observed scale $R$, and that observations corresponding to different $R$ originate at different scale factors, $a$.
During matter domination, where $H\approx H_{0}\sqrt{\Omega_{m}}a^{-3/2}$, for sub-horizon modes
\begin{equation}\label{eq:subhor}
\frac{k}{H_0}\gtrsim\sqrt{\frac{\Omega_m}{a}}.
\end{equation}
As noted previously, for a given scale $R$, a physical fluctuation $\delta_m(R)$ receives contributions from Fourier modes $\delta_k$ where $k\lesssim 1/R$. If a sub-horizon instability is present for all $a<a_i$, it will affect all modes with a physical scale today of 
\begin{equation}
R< \frac{1}{H_0}\sqrt{\frac{a_i}{H_0}}\approx 5500\sqrt{a_i}\,\frac{\rm Mpc}{h}.
\end{equation}
Assuming sub-horizon instabilities for $a<a_i$, we make the conservative assumption that all $\delta_m(R)$ with $R<(H_i a_i)^{-1}$ will have a minimal value today of $\delta_{\rm min}(a_i)$ since the fluctuation will receive contributions from at least one unstable Fourier mode $\delta_k$. We can thus rule out all values of the scale factor $a_i$, for which $\delta_m(R)<\delta_{\rm min}(a_i)$, where $k=R^{-1}$ represents the Fourier mode with the smallest wavelength that contributes to a fluctuation at scale $R$. This constrains the observationally viable value of $a_i$ to 
\begin{equation}\label{eq:qAineq}
a_i<\left(\frac{\delta_m(R)q\Omega_m}{2}\right)^{1/2}.
\end{equation}
Since the observed $\delta_m(R)$ is smallest at large scales, we expect large scales to give the most stringent constraints on $a_i$. However, if the Fourier modes contributing to the fluctuation are outside the horizon, they will not be affected by the instability. From Eq.~\ref{eq:subhor}, this corresponds to
\begin{equation}\label{eq:qAineq2}
a_i<\left[\frac{2\cdot 10^{-4}}{k}\left(\frac{{\rm Mpc}}{h}\right)^{-1}\right]^2.
\end{equation}
We thus have two inequalities for $a_i(R=k^{-1})$, Eqns.~\ref{eq:qAineq} and \ref{eq:qAineq2}, for which if any is fulfilled, instabilities will not cause disagreement with current observations.  

\subsection{The $\beta_1$-model}\label{sec:b1}

Since we have assumed that the background geometry is flat, it is not possible to self-consistently derive limits with parameter values (in this case, $\beta_i$) corresponding to valid background solutions. However, regarding the asymptotic flat potential assumption as just a way of saying that a fluctuation should have no effect on the asymptotic space time behaviour, we can study our solutions using parameter values that give non-flat backgrounds. In the case of only $\beta_1$, and $m\sim H_0$, in the de Sitter limit this corresponds to $\beta_1=c^{-1}=\sqrt{3}$ and $q=1/3$.  However, this value
depends on the metric ansatz Eq.~\ref{eq:metricansatz}. Using instead the metric ansatz of \rcite{2013JHEP...10..031E}, the corresponding value is $q=28/3$, that is, the Vainshtein radius differs by a factor of $\sim 3$. In the following, we will therefore account for how the derived limits scale with $q$, and keep in mind the potential uncertainty of the limits due to possible changes in the value of $q$. 

\begin{figure}
\begin{centering}
\includegraphics[scale=0.65]{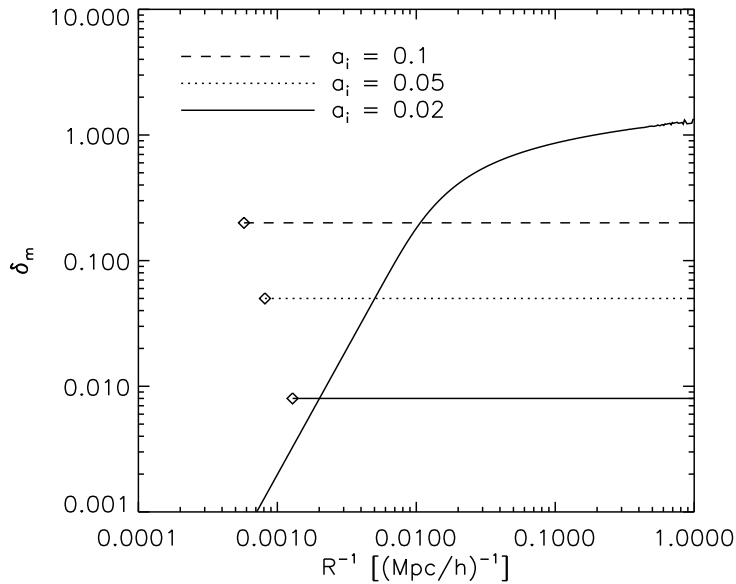}
\par\end{centering}
\caption{\label{fig:delta3} A comparison between the GR prediction for $\delta_m(R)$ (solid curved line) and the minimum  value $\delta_{\rm min}(a_i)$ today for an instability present at $a<a_i$ (horizontal lines) for the $\beta_1$-model using $q=1/3$. The range of $k=R^{-1}$ for each line (indicated by the diamond) corresponds to scales that are sub-horizon for any $a<a_i$, and thus are affected by the instability.}
\end{figure}

In Fig.~\ref{fig:delta3}, we plot the GR prediction for $\delta_m(R)$ as outlined in Sec.~\ref{sec:sf}, for the $\beta_1$-model using $q=1/3$. Note that
a physical fluctuation $\delta_m(R)$ will receive contributions from all $\delta_k$ where $k\lesssim 1/R$ (Eq.~\ref{eq:gaussfilter}) and that $\delta_m\lesssim 1$ while most $\delta_k\gtrsim 1$. 
The horizontal lines correspond to the minimum perturbation amplitude today $\delta_{\rm min}(a_i)$, for an instability present at $a<a_i$ for the $\beta_1$-model. The range of $k=R^{-1}$ for each line (indicated by the diamond) corresponds to scales that are sub-horizon for any $a<a_i$. It is evident that the larger the  $a_i$, the more dramatic the modifications to the observed structures since it will impose a larger minimum value of $\delta_m$ over a larger range of physical scales. 

\begin{figure}
\begin{centering}
\includegraphics[scale=0.65]{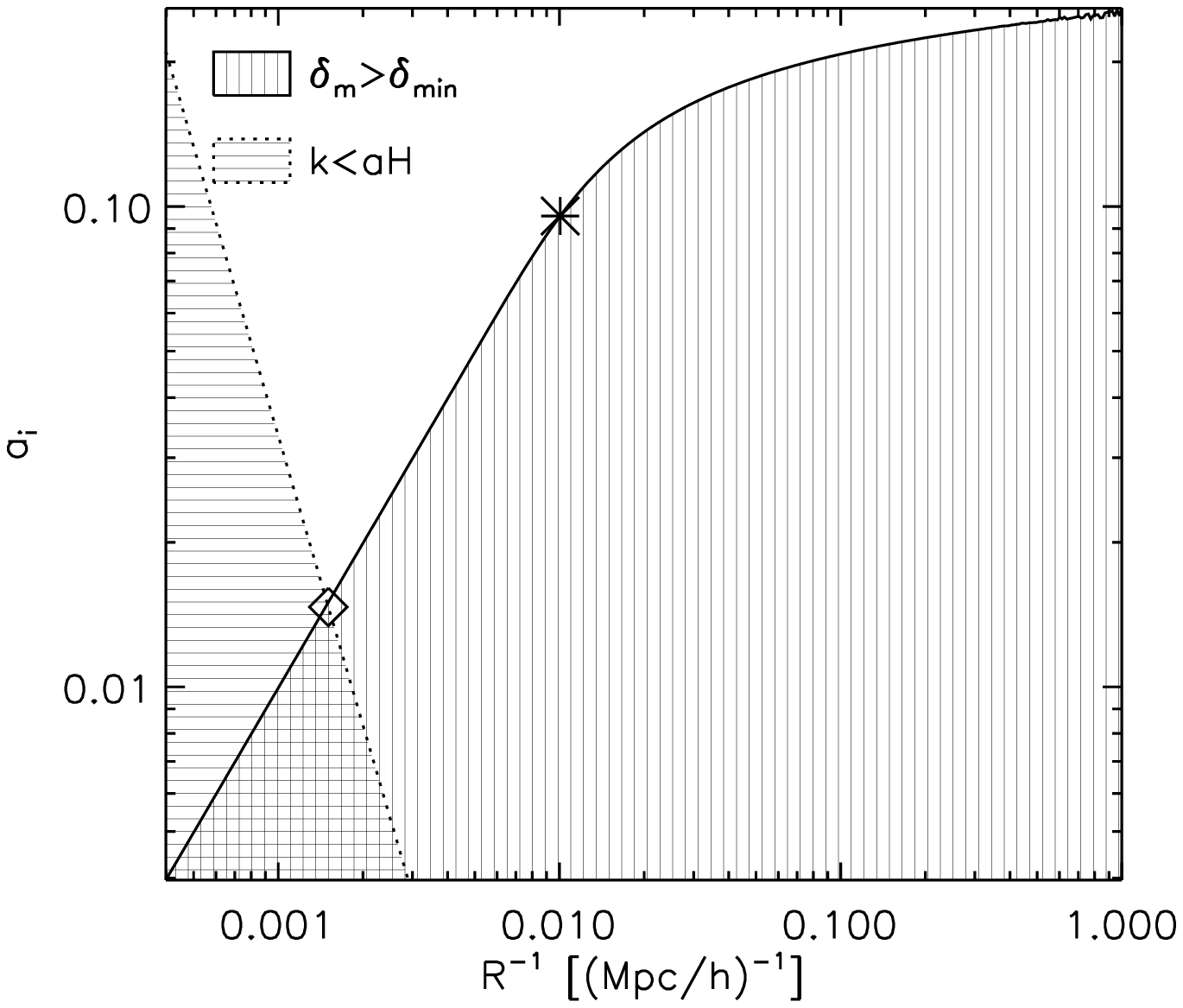}
\par\end{centering}
\caption{\label{fig:delta2} Constraints in the $\beta_1$-model with $q=1/3$ on the maximum value of $a_i$ up to which sub-horizon gradient instabilities are effective, as a function of the observational scale $R$. The hatched regions show regions {\em not} excluded by observations, since either the observed $\delta_m$ is larger than $\delta_{\rm min}$ (Eq.~\ref{eq:qAineq}), or all Fourier modes contributing to the fluctuation were outside the horizon when the instability was active (Eq.~\ref{eq:qAineq2}). If any of these constraints is fulfilled, the model is in agreement with current observations. The white (non-hatched) region thus represents values of $a_i$ in principle excluded. The diamond indicates the minimum allowed value of $a_i\approx 0.015$, corresponding to $z_i\approx 70$ from observations at $R\sim 0.7\,{\rm Gpc}/h$. However, since low redshift observations currently only probe $R\lesssim 100\,{\rm Mpc}/h$, the current constraint indicated by the asterisk is $a_i\lesssim 0.1$, or $z_i\gtrsim 9$.}
\end{figure}

In Fig.~\ref{fig:delta2}, constraints given by Eqns.~\ref{eq:qAineq} and \ref{eq:qAineq2} for the $\beta_1$-model with $q=1/3$ are shown. Hatched regions indicate values {\em not} excluded by observations, since either the observed $\delta_m$ is larger than $\delta_{\rm min}$, or all Fourier modes contributing to the fluctuation were outside the horizon when the instability was active.
Note that this is not a constraint plot for $a_i$ and $R^{-1}$, but rather for $a_i$ as a function of $R^{-1}$.  We can read off a minimum $a_i\approx 0.015$, corresponding to $z_i\approx 70$ from observations at $R^{-1}\approx 0.0015\,({\rm Mpc}/h)^{-1}$, or $R\sim 0.7\,{\rm Gpc}/h$, indicated by the diamond symbol.

However, $R^{-1}\sim 0.0015\,({\rm Mpc}/h)^{-1}$ corresponds to $k/H_0\sim 4.5$. These modes are only probed by CMB observations originating at $z\sim 1100$, thus only probing instabilities at $a\lesssim 10^{-3}$, unless secondary CMB constraints such as gravitational lensing and the integrated Sachs-Wolfe (ISW) effect are important. Primary CMB observations do not constrain instabilities at $a\gtrsim 10^{-3}$. Low redshift galaxy observations on the other hand probe modes with $R^{-1}\gtrsim 0.01\,({\rm Mpc}/h)^{-1}$ and constrain $a_i\lesssim 0.1$, or $z_i\gtrsim 9$ for the $\beta_1$-model with $q=1/3$, indicated by the asterisk in Fig.~\ref{fig:delta2}. Since $a_i\propto q^{1/2}$ (Eq.~\ref{eq:qAineq}), changing the value of $q$, will change the constraint on $a_i$. Low redshift observations thus rule out 
sub-horizon gradient instabilities at a scale factor larger than $a_i\approx 0.15q^{1/2}$, or redshift smaller than $z_i\approx 6.7/q^{1/2}-1$. For example, using instead the value $q=28/3$ as derived using the metric ansatz of \rcite{2013JHEP...10..031E}, the corresponding limit for the $\beta_1$-model is $a_i\lesssim 0.5$, or $z_i\gtrsim 1$.

We note also that as $c\to 0$ or $c\to\infty$, it is possible that $q\to 0$ or $q\to\infty$, depending on the values of the $\beta_i$. As $q\rightarrow\infty$, the instabilities will be pushed into the far future, since $\rho_V \rightarrow 0$.

\section{Summary and discussion}
In this paper, we have investigated the observational consequences of sub-horizon linear gradient instabilities under the assumption that the Vainshtein mechanism restores GR at scales when the linear approximation breaks down. For static, spherically symmetric matter distributions, the linear approximation breaks down for density fluctuations $\rho_m>\rho_V$, where $\rho_V\sim \rho^0_{\rm crit}$ for $m\approx H_0$. If the Vainshtein mechanism operates at similar scales also for quasi-static, non-spherical matter fluctuations, we have ruled out sub-horizon gradient instabilities at scale factors $a_i\gtrsim 0.15q^{1/2}$, or in terms of redshift, $z_i\lesssim 6.7/q^{1/2}-1$. Here, $q=q(\beta_i,c)$ that for $[\beta_i,c]\approx 1$ is of order unity, but in principle can take any value between zero and infinity. For the $\beta_1$-model with all $\beta_i=0$ except $\beta_1$, depending on the assumed value of $q$, we can constrain $z_i\gtrsim 1-10$. This is to be compared to theoretical predictions for the same model, showing that the instability will persist down to a redshift of $z_i\approx 0.5$ \cite{Konnig:2014dna}. At face value, this indicates a serious tension between observational constraints and theoretical predictions for the $\beta_1$-model. However, the difference is sufficiently small that the final verdict on the $\beta_1$-model needs to await a more detailed analysis. For cases where the theoretical prediction and the observational constraints differ by larger amounts, the method outlined in this paper should be sufficient to rule in or out the models, in terms of the effects of gradient instabilities. 

Since the largest observable scales are the most efficient in constraining the effect of instabilities, we note that a study of secondary effects on the CMB, particularly the integrated Sachs-Wolfe (ISW) effect, could potentially yield more stringent constraints on models. The ISW effect probes time variations in the gravitational potentials between the CMB decoupling and the present epoch. In a matter dominated universe, we expect potentials to be constant and the ISW effect to be zero. If the effect from curvature and/or dark energy on the background expansion is non-negligible, potentials will decay with time and we will observe slightly higher CMB temperatures in the direction of matter over densities. This effect has been detected using the angular cross-power spectrum between the CMB temperature anisotropies and large-scale structure at around $3\sigma$ significance, see \cite{Enander:2015vja} for a list of references and a study of the observational effects on a bimetric model not affected by gradient instabilities.
If density fluctuations would grow exponentially, so would the potentials. The potential will only vary in time exponentially during a very short epoch directly after entering inside the horizon, after which it will grow as $\Phi\propto a^2$. Since what is observed is decaying potentials, ISW observation can potentially put stringent constraints on models plagued by gradient instabilities. It also remains to be investigated to what extent the qualitative analysis presented here can also be applied to the tensor sector. 

\acknowledgments

We would like to thank Yashar Akrami, Adam R. Solomon, Frank K\"{o}nnig,  
Fawad Hassan and Angnis Schmidt-May
for many interesting discussions in relation to this work. E.M. acknowledges support for this study by the Swedish Research Council.

\bibliographystyle{JHEP}
\bibliography{bibliography}{}

\end{document}